\documentclass{moriond}

\usepackage{amsmath}
\usepackage{xcolor}
\usepackage{subcaption}

\def\be{\begin{equation}}
\def\ee{\end{equation}}
\def\bea{\begin{eqnarray}}
\def\eea{\end{eqnarray}}

\newcommand{\Photo}{\includegraphics[height=45mm]{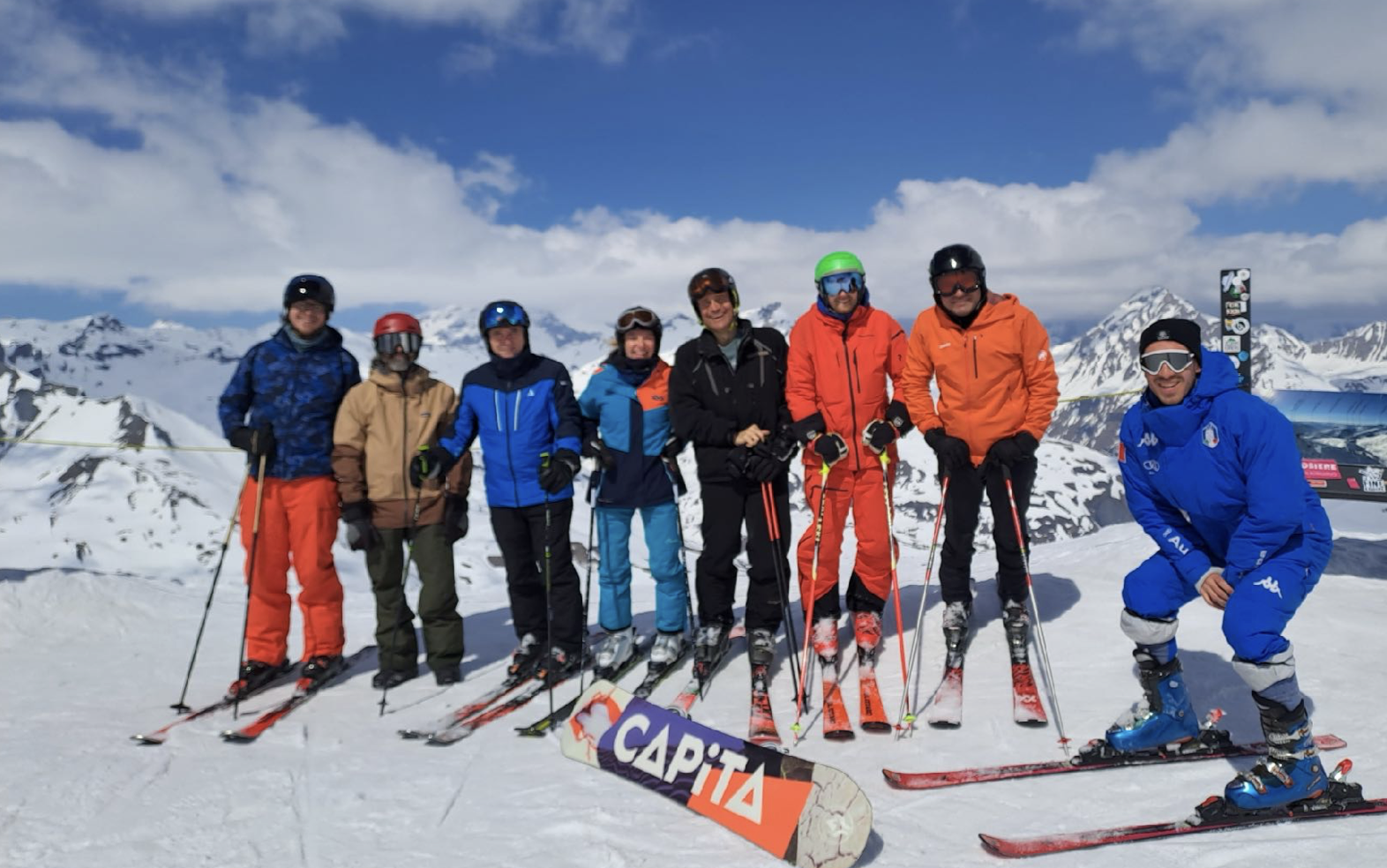}}

\begin{document}
\vspace*{2.0cm}
\title{Infrared singularities and the collinear limits of multi-leg scattering amplitudes}

\author{ Sebastian Jaskiewicz }

\address{Albert Einstein Center for Fundamental Physics,
Institut f\"ur Theoretische Physik,\\ Universit\"at Bern,
Sidlerstrasse 5, CH-3012 Bern, Switzerland }

\maketitle\abstracts{
Scattering amplitudes admit a factorised structure in special kinematic limits, such as the soft and collinear limits. In this work, we investigate the multi-particle collinear limits of massless amplitudes at high perturbative orders, focusing on the exploration of the mechanisms via which strict collinear factorisation of $n$-particle scattering amplitudes is realised when $m$ particles become collinear. We show through four loops that the requirements on the structure of the massless soft anomalous dimension that are imposed by strict collinear factorisation in all two-particle collinear limits are enough to guarantee factorisation also in any multi-particle collinear limit. Demanding that strict collinear factorisation of massless partons is satisfied also for amplitudes that contain a massive coloured particle, we derive new constraints on the soft anomalous dimension by considering the collinear limit of three massless particles. 
}

\vspace{-1.2cm}

\section{Introduction}
\label{sec:Introduction}
\vspace{-0.2cm}

Investigations of scattering amplitudes in various kinematic limits are an 
attractive prospect since in these specific configurations 
scattering amplitudes factorise, offering insights into the structure of gauge theories. 
A particularly rich kinematic limit is the collinear limit 
of massless partons. The expectation is such that the collinear sector will factorise 
from the rest of the scattering process, which can yield information on the form of the
allowed functions that describe the process. While two-particle collinear limits were considered previously, this proceeding is based on a recent work
investigating the multi-collinear limits of scattering amplitudes~\cite{Duhr:2025cye}. 

\vspace{-0.4cm}

\section{Soft anomalous dimension}
\vspace{-0.2cm}
We begin our considerations with a scattering amplitude ${\mathcal{M}}_n$ for $n$ massless coloured partons. Scattering amplitudes develop infrared divergences that are known to factorise. Namely, they can be contained in a \emph{universal} factor ${{\bf Z}}_n$ which multiplies a finite, in the $\epsilon \to 0$ limit, hard function,~${\cal H}_n$.  The $n$-point UV renormalised amplitude ${\mathcal{M}}_n$ has the following structure \cite{Becher:2009qa,Gardi:2009zv,Agarwal:2021ais}
\begin{equation}
\label{eq:IRfacteq}
{\mathcal{M}}_n \left(\{p_i\},\mu, {\alpha_s} (\mu^2),\epsilon \right) \, = \, 
{\bf Z}_n \left(\{p_i\},\mu_f, {\alpha_s} (\mu_f^2), \epsilon \right)
{\mathcal{H}}_n \left(\{p_i\},\mu_f, \mu, {\alpha_s} (\mu^2) \right).
\end{equation}
Here, $\{p_i\}$ are the momenta of external particles, $\mu$ is the UV renormalisation scale, and $\mu_f$ is the scale at which IR singularities are regularised. The renormalisation 
factor ${\bf Z}_n$ obeys an RG equation, the solution to which can
be  expressed in terms of a  path-ordered exponential of the  
\emph{soft anomalous dimension} ${\bf \Gamma}_n$ as follows
\begin{equation} 
\label{eq:RGsol}
{\bf Z}_n \left(\{p_i\},\mu_f, {\alpha_s} (\mu_f^2),\epsilon  \right) \, = \,  
{\cal P} \exp \left\{ -\frac{1}{2}\int_0^{\mu_f^2} \frac{d \lambda^2}{\lambda^2}\,
{\bf \Gamma}_n \left(\{p_i\},\lambda, {\alpha_s}(\lambda^2) \right) \right\}\,. 
\end{equation}
The massless soft anomalous dimension has been fully determined to 
three loop order~\cite{Almelid:2015jia} and its structure was investigated at four loops~\cite{Becher:2019avh}. 
We write 
${\bf \Gamma}_n$ as~\cite{Falcioni:2021buo} 
\begin{eqnarray}\label{eq:adm-param}
    {\bf{ \Gamma}}_{n}\left(\{s_{ij}\},\lambda,\alpha_s(\lambda^2) \right)
     &=& {\bf{ \Gamma}}^{{\rm{dip.}}}_{n}\left(\{s_{ij}\},\lambda,\alpha_s  \right)
     + {\bf{ \Gamma}}_{n,4{\rm{T}}-3{\rm{L}}}(\alpha_s)
     + {\bf{ \Gamma}}_{n,4{\rm{T}}-4{\rm{L}}}(\{\beta_{ijkl}\},\alpha_s)
    \nonumber\\ && 
    + {\bf{ \Gamma}}_{n,{\rm{Q}}4{\rm{T}}-2,3{\rm{L}}}(\{s_{ij} \},\lambda,\alpha_s)  
    + {\bf{ \Gamma}}_{n,{\rm{Q}}4{\rm{T}}-4{\rm{L}}}(\{\beta_{ijkl} \} ,\alpha_s)  
    \\ \nonumber && 
    + {\bf{ \Gamma}}_{n,5{\rm{T}}-4{\rm{L}}}(\{\beta_{ijkl} \} ,\alpha_s)  
        + {\bf{ \Gamma}}_{n,5{\rm{T}}-5{\rm{L}}}(\{\beta_{ijkl} \} ,\alpha_s)
        + \mathcal{O}(\alpha_s^5)\,.
\end{eqnarray}
${\bf{ \Gamma}}^{{\rm{dip.}}}_{n}$ is the dipole formula, which encodes pairwise interactions between the partons. It constitutes the full result for massless 
soft anomalous dimension up to the two loop order. The next two terms, ${\bf{ \Gamma}}_{n,4{\rm{T}}-3{\rm{L}}}$ and
 ${\bf{ \Gamma}}_{n,4{\rm{T}}-4{\rm{L}}}$, encode 
the correction to the dipole formula which first arises at the three loop order~\cite{Almelid:2015jia}, and the terms in the last two lines are the 
corrections which start at four loops~\cite{Becher:2019avh,Henn:2019swt}. The explicit forms of these
terms are rather lengthy, but can be found in a suitable form for this discussion in Sec.~2 of the publication on which this proceeding is based~\cite{Duhr:2025cye}.

The soft anomalous dimension depends on kinematic invariants 
formed between the external partons and \emph{conformally invariant cross-ratios} (CICRs)~\cite{Becher:2009qa,Gardi:2009qi} (the $\beta_{ijkl}=\ln\rho_{ijkl}$)
\begin{eqnarray}\label{eq:s_ij}
(-s_{ij}) = 2 |p_i \cdot p_j | e^{-i \pi \lambda_{ij}},
\hspace{2cm}
\rho_{ijkl} = \frac{(-s_{ij})(-s_{kl})}{(-s_{ik})(-s_{jl})} ,
\end{eqnarray}
where $\lambda_{ij} = 1$ if the partons $i$ and $j$ are both in either the initial or the final state, and $\lambda_{ij} = 0$ otherwise. If an external massive parton is present, we need to add terms proportional to $F_{{\rm{h}}2}\left(r_{ijI} \right)$ and $F_{{\rm{h}}3}\left(r_{ijI},r_{ikI},r_{jkI}\right)$  to  eq.~\eqref{eq:adm-param}~\cite{Liu:2022elt}.
Results for these are known at three loops from \cite{Liu:2022elt} and~\cite{Gardi:2025lws}, respectively. Variables $r_{ijI}$ are massive versions of the CICRs: $r_{ijI} = {p_i\cdot p_j\, p_I^2}/{(2 p_i\cdot p_I\, p_j\cdot p_I )}$.

\vspace{-0.4cm}

\section{Multi-collinear limits of scattering amplitudes}

\vspace{-0.2cm}

\begin{figure}
\begin{center}
\begin{center}
\includegraphics[width=0.27\textwidth]{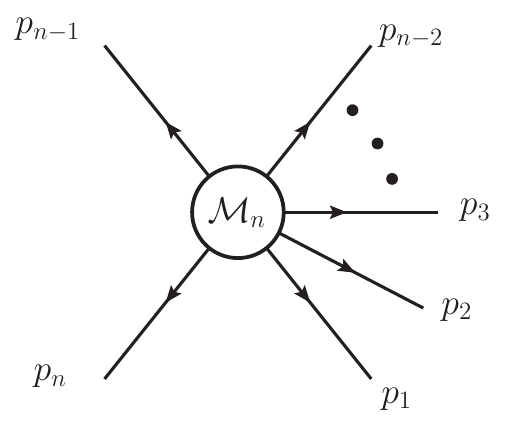} 
\includegraphics[width=0.27\textwidth]{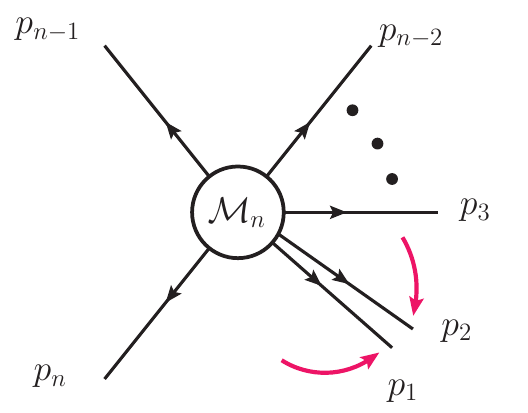} 
\includegraphics[width=0.27\textwidth]{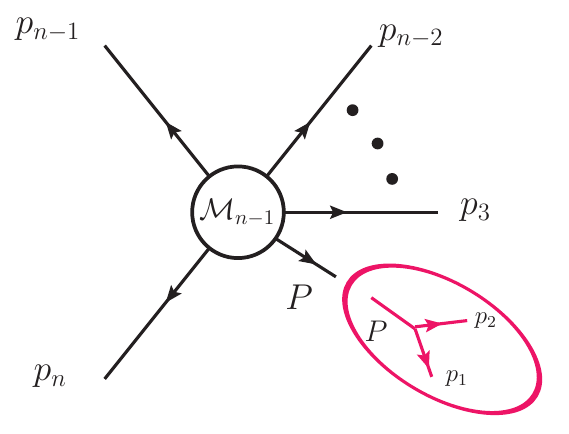}
\end{center}
\caption{\label{fig:collinearlimit} Diagrams depicting the collinear limit approach from an $n$-point scattering amplitude. The object circled in red represents
the splitting amplitude which is factorised from the $n-1$-point amplitude and depends only on the degrees of freedom of the collinear particles.}
\end{center}
\end{figure}

\begin{figure}[t]
\begin{center}
    \includegraphics[width=0.40\textwidth]{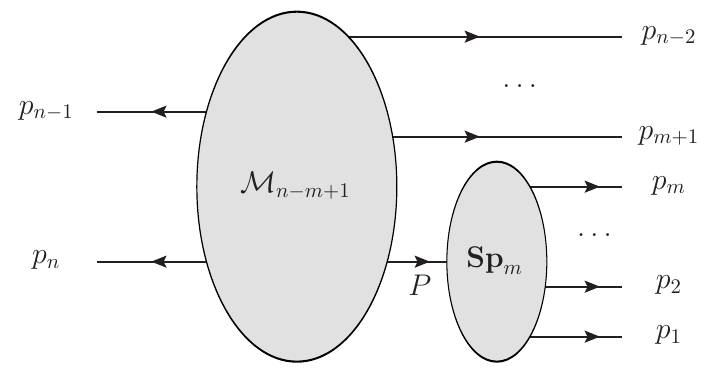}
\end{center}
    \caption{Diagram showing the multi-particle timelike 
        collinear limit of an $n$-point scattering amplitude considered in this work. }
\label{fig:m-collinearTL}
\end{figure}

The central topic of this
work is concerning the multi-collinear limits of $n$-point 
scattering amplitudes. This kinematic limit is characterised by
allowing for a subset of invariants 
$p_i\cdot p_j$ to become parametrically 
small, as shown in Fig.~\ref{fig:collinearlimit}.  
In the collinear limit, the QCD amplitude develops an additional singularity due to the parent parton being close to on-shell.  We investigate the factorisation properties of ${\mathcal{M}}_n $ in this limit. In particular, we focus on the \emph{timelike} $m$-parton collinear 
limit. Here, the $m$ collinear partons are all either in the initial 
or all in the final state. See Fig.~\ref{fig:m-collinearTL} for the latter case. 
In this limit, the $n$-point amplitude ${\mathcal{M}}_n(p_1,\ldots p_n;\mu)$  factorises into an $(n-m+1)$-point amplitude multiplied by a  splitting amplitude, which we denote by~${{\bf Sp}}_{m}$. The splitting amplitude captures the divergence at $P^2= 0$ and depends only on the momenta and colour of the $m$ collinear 
particles 
\cite{Berends:1988zn,Mangano:1990by,Bern:1995ix,Kosower:1999xi}
\begin{eqnarray}
\label{TL-collinear-limit} 
{\mathcal{M}}_n(p_1,\ldots p_m, \{p_i\}_{\text{rest}};\mu) 
&\stackrel{p_1 \parallel p_2 \parallel \ldots \parallel p_m}{\longrightarrow}&
{{\bf Sp}}_{m}(p_1, \ldots p_m;\mu) \,{\mathcal{M}}_{n-m+1}(P,\{p_i\}_{\text{rest}};\mu)\,,    
\end{eqnarray}
where $P$ is the total momentum of the collinear particles, $P= p_1 + \ldots + p_m$. Due to the factorisation structure in eq.~\eqref{TL-collinear-limit}, the IR singularities 
of the timelike ${{\bf Sp}}_{m}$ are given in terms of the so-called \emph{splitting amplitude 
soft anomalous dimension} (${\bf \Gamma}_{{{\bf Sp}},m}$), which is related to the soft anomalous dimensions of the $n$- and $(n-m+1)$-point amplitudes~\cite{Duhr:2025cye,Becher:2009qa,Dixon:2009ur,Catani:2011st}. Namely, 
\begin{align}\label{GammaSPdef} 
\begin{split}
{\bf \Gamma}_{{{\bf Sp}},m} (p_1,\ldots p_m;\mu_f)
&={\bf \Gamma}_n (p_1,\ldots p_m, p_{m+1},\ldots p_n;\mu_f) 
\\&\hspace*{30pt} -\,{\bf \Gamma}_{n-m+1}(P, p_{m+1},
\ldots p_n;\mu_f)|_{{{\bf T}}_P\to \sum_{i=1}^m {{\bf T}}_i}\,,
\end{split}
\end{align}
where ${{\bf T}}_P$ is the colour charge of the collinear particles, ${{\bf T}}_P = {{\bf T}}_1 + \ldots + {{\bf T}}_m$. 
Eq.~\eqref{GammaSPdef} is central to our work. The reason for this is manifest in the structure, that is, while the right-hand side of eq.~\eqref{GammaSPdef} contains dependence on all partons present in the scattering, the left-hand side contains information only on the degrees of freedom of the partons becoming collinear. This is known as \emph{strict-collinear factorisation}~\cite{Catani:2011st}, and can easily be shown to hold up to two-loops using the dipole formula, the ${\bf{ \Gamma}}^{{\rm{dip.}}}_{n}$ in eq.~\eqref{eq:adm-param}. The two-particle collinear limit at three and four loops has already been used to constrain the form of ${\bf \Gamma}_n$ given that strict-collinear factorisation holds~\cite{Almelid:2015jia,Becher:2019avh,Almelid:2017qju}.

To investigate this remarkable property at high-loop orders and for multi-collinear limits, we compute and study the various contributions to ${\bf \Gamma}_{{{\bf Sp}},m}$ by directly inserting terms from the soft anomalous dimension, given schematically in eq.~\eqref{eq:adm-param}, into eq.~\eqref{GammaSPdef}. The calculations are lengthy and are not presented here, however detailed derivation can be found in the parent publication~\cite{Duhr:2025cye}. Given the expectation from strict-collinear factorisation described above, there are two complementary viewpoints that can be taken in these calculations. Firstly, for the cases where the soft anomalous dimension is known, the factorisation properties of the scattering amplitude in the multi-collinear limits can be \emph{tested}. Secondly, assuming that the result for the soft anomalous dimension is not known and demanding strict-collinear factorisation holds, we can derive constraints on the form of terms in the soft anomalous dimension which can aid in a \emph{bootstrap} program~\cite{Almelid:2017qju}.
Here, we state the main conclusions of this work~\cite{Duhr:2025cye}: 
\begin{itemize}
    \item The two-particle collinear limit constraints used in previous works~\cite{Almelid:2015jia,Almelid:2017qju} have been obtained by considering differences between anomalous dimensions with fixed $n$. For example, demanding that ${\bf \Gamma}_3$ is equal to ${\bf \Gamma}_4-{\bf \Gamma}_3$ leads to constraints on the soft anomalous dimension. In this work, we have shown the universality of this constraint, namely the fact that no new constraint information is gained by increasing the number of external legs and taking instead ${\bf \Gamma}_n-{\bf \Gamma}_{n-1}$. We have shown the two-particle constraint to hold for ${\bf \Gamma}_n$ with any $n$. 
    \item We have also shown that for fully massless scattering amplitudes, the intricate interplay between colour and kinematics ensures multi-collinear limits are directly satisfied at three and four loops, as soon as all two-particle collinear limits are satisfied. Therefore, no new information for the bootstrap program arises from investigation of multi-collinear limit of massless amplitudes on top of the two-particle collinear limit up to the fourth loop order. 
    \item For the case where a massive coloured parton is present in the amplitude, we have shown that a three-particle collinear limit does indeed lead to a new constraint on the soft anomalous dimension terms on top of the two-particle collinear limit: 
    \begin{eqnarray}\label{eq:M-3pc-constraint}
 \lim_{p_a||p_b||p_c} 
F_{{\rm{h}}3}\left(r_{abI},r_{acI},r_{bcI}  \right) 
 =
\lim_{r_{bcI}\to 0} F_{{\rm{h}}3}\left( \beta_{abIc}, \beta_{acIb} \,;r_{bcI} \right)   = {4}{\cal F} (\beta_{ablc},\beta_{aclb})\Big\rvert_{p_a||p_b||p_c}\,,
\end{eqnarray}
where ${\cal F}$ are the massless quadruple kinematic terms in ${\bf{ \Gamma}}_{n,4{\rm{T}}-4{\rm{L}}}$. This limit agrees with the small-mass limit \cite{Duhr:2025cye} computed in~\cite{Liu:2022elt}.
\end{itemize}

\vspace{-0.75cm}

\section{Concluding remarks and outlook}
\vspace{-0.25cm}

In this work, we have systematically investigated the timelike multi-collinear limits
of scattering amplitudes through four loops. From the point of view of a bootstrap program, 
in the massless case we find that demanding strict collinear factorisation of multi-collinear limits does not lead to additional constraints on the form of the soft anomalous dimension, on top of ones obtained from two-particle collinear limits. 
On the other hand, if we consider an amplitude containing additionally one massive coloured parton, then the multi-collinear limits do indeed provide another constraint~\cite{Duhr:2025cye}. Here, we have focused on the timelike limit. However, another rich collinear limit is the spacelike configuration where plenty of recent developments regarding factorisation breaking and appearance of coherence violating logarithms occurred~\cite{Catani:2011st,Forshaw:2012bi,Schwartz:2017nmr,Cieri:2024ytf,Duhr:2025lyg,Henn:2024qjq,Buccioni:2026mfg,Becher:2024kmk,Becher:2025igg,Nabeebaccus:2023rzr,Banfi:2025mra,Dasgupta:2025cgl,Becher:2026kbr,Chen:2026dnj,Barcaro:2026dsd}. It would be interesting to utilise the technology developed here to gain insights also into this limit.

\vspace{-0.4cm}

\section*{Acknowledgments}
\vspace{-0.2cm}
I would like to thank C. Duhr, E. Gardi, J. L\"ubken, and L. Vernazza for collaboration and E.~Gardi for useful comments. 
This work has been supported by the Swiss National Science Foundation Ambizione grant PZ00P2\_223524.

\vspace{-0.40cm}

\section*{References}

\vspace{-0.3cm}

\end{document}